\documentclass{llncs}
   \usepackage{times}
   \usepackage{mathptm}
   \usepackage{epsfig}
   \usepackage{color}
   \usepackage{latexsym}
   \usepackage{pslatex}

\begin{document}

%\title{Speeding-Up NLFSR-Based Stream Ciphers by Feedback Function Transformation}
\title{An Equivalence Preserving Transformation from \\ the Fibonacci to the Galois NLFSRs}
\author{Elena Dubrova}
\institute{Royal Institute of Technology (KTH), Electrum 229, 164 46 Kista, Sweden\\
\email{dubrova@kth.se}}

\maketitle

\begin{abstract}
Conventional Non-Linear Feedback Shift Registers (NLFSRs) use the
Fibonacci configuration in which the value of the first bit is updated
according to some non-linear feedback function of previous values of
other bits, and each remaining bit repeats the value of its previous
bit.  We show how to transform the feedback function of a Fibonacci
NLFSR into several smaller feedback functions of individual bits. Such
a transformation reduces the propagation time, thus increasing the
speed of pseudo-random sequence generation. The practical significance
of the presented technique is that is makes possible increasing the
keystream generation speed of any Fibonacci NLFSR-based stream cipher
with no penalty in area.
\end{abstract}


\section{Introduction}

Non-Linear Feedback Shift Registers (NLFSRs) have been proposed as an
alternative to Linear Feedback Shift Registers (LFSRs) for generating
pseudo-random sequences for stream ciphers. NLFSR-based stream ciphers
include Achterbahn~\cite{GaGK07}, Dragon~\cite{CHM05}, Grain~\cite{hell-grain},
Trivium~\cite{canniere-trivium}, VEST~\cite{cryptoeprint:2005:415},
and~\cite{GaGK06}. NLFSRs have been shown to be more resistant to
cryptanalytic attacks than LFSRs~\cite{1241371,Ca05}.  However,
construction of large NLFSRs with guaranteed long periods remains an
open problem. A systematic algorithm for NLFSR synthesis has not been
discovered so far. Only some special cases have been
considered~\cite{Golomb_book,Mykkeltveit77,MykkeltveitST79,Ro84,Ja89,Ro92,603256,838191,JaS04}.

\begin{figure}[t!]
\begin{center}
\resizebox{0.75\columnwidth}{!} {\input{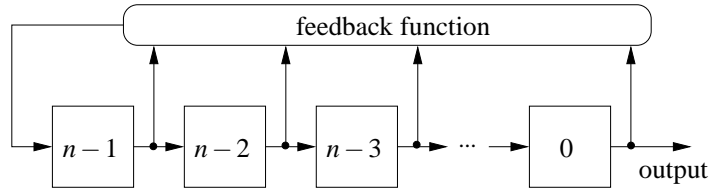}}
\caption{An Fibonacci type of NLFSR.}\label{nlfsr_f}
\end{center}
\end{figure}

In general, there are two ways to implement an NLFSR: in the Fibonacci
configuration, or in the Galois configuration.  The {\em Fibonacci}
configuration, shown in Figure~\ref{nlfsr_f}, is conceptually more
simple. The Fibonacci type of NLFSRs consists of a number of bits
numbered from left to right as $n-1, n-2, \ldots, 0$ with feedback
from each bit to the $n-1$th bit. At each clocking instance, the value
of the bit $i$ is moved to the bit $i-1$. The value of the bit 0
becomes the output of the register. The new value of the bit $n-1$ is
computed as some non-linear function of the previous values of other bits.

\begin{figure}[t!]
\begin{center}
\resizebox{0.99\columnwidth}{!} {\input{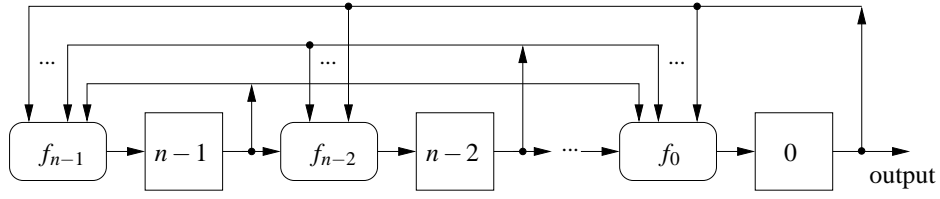}}
\caption{A Galois type of NLFSR.}\label{nlfsr_g}
\end{center}
\end{figure}

In the {\em Galois} type of NLFSR, shown in Figure~\ref{nlfsr_g}, each
bit $i$ is updated according to its own feedback function. Thus, in
contrast to the Fibonacci NLFSRs in which feedback is applied to the
$n-1$th bit only, in the Galois NLFSRs feedback is potentially applied
to every bit. Since the next state functions of individual bits of a
Galois NLFSR are computed in parallel, the propagation time is reduced
to that of smaller functions of individual bits. This makes Galois
NLFSRs particularly attractive for stream ciphers application in which
high keystream generation speed is important.

However, Galois NLFSRs also have the following two drawbacks:
\begin{enumerate}
\item An $n$-bit Galois NLFSR with the period of $2^n-1$ does not 
necessarily satisfy the 1st and the 2nd postulates of
Golomb~\cite{DuTT08}. An $n$-bit Fibonacci NLFSR with the period of
$2^n-1$ always satisfy both postulates~\cite{Golomb_book}.
\item The period of the output sequence of a Galois NLFSR is not 
necessarily equal to the length of the longest cyclic sequence of its
consecutive states~\cite{DuTT08}. The period of a Fibonacci NLFSR always
equals to the longest cyclic sequence of its
consecutive states~\cite{Golomb_book}.
\end{enumerate}

\begin{figure}[t!]
\begin{center}
\resizebox{0.9\columnwidth}{!} {\input{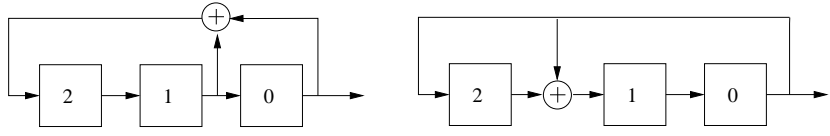}}
\caption{The Fibonacci LFSR (left) and the Galois LFSR (right) for the generator polynomial $x^3+x+1$.}\label{lfsr}
\end{center}
\end{figure}

These drawbacks do not create any problems in the linear
case because, for LFSRs, there exist a one-to-one mapping between 
the Fibonacci and Galois configurations. A Galois LFSR generating the
same output sequence as a given Fibonacci LFSR (and therefore
possessing none of the above mentioned drawbacks) can be obtained by
reversing the order of the feedback taps and adjusting the initial
state. For example, Figure~\ref{lfsr} shows the
Fibonacci and Galois configurations for the generator polynomial $x^3+x+1$. If
the Fibonacci LFSR is initialized to the state $001$ and the Galois
one is initialized to the state $101$, then they generate the same
periodic sequence $1001011$.

In the non-linear case, however, no mapping between the Fibonacci and
the Galois configurations has been known until now. The problem of
finding such a mapping is addressed in this paper.
%A natuar question to ask is the following: Is is possible to dene a
%subset of Galois NLFSRs for which a mapping from the Fibonacci NLFSRs
%exists? If such a subset can be specified, and if the mapping
%is easy to do, then we can first construct a
%Fibonacci NLFSR and then transform it to its Galois equivalent. The
%resulting NLFSR will posses the speed advantage of Galois type, but
%will inherit none of the disadvantages.
%Content
We show that, for each Fibonacci NLFSR, there exist a class of
equivalent Galois NLFSRs which produce the same output sequence. 
%These
%NLFSRs provide an advantage in speed and do not posess the above
%mentioned drawbacks. 
We show how to transform a given Fibonacci NLFSR into an equivalent
Galois NLFSR. 
%We also identify the unique representative from the
%class of equivalent NLFSRs which minimizes the number of feedback
%variables, and give an algorithm for its construction.

%Since a Fibonacci NLFSR can have many equivalent Galois NLFSRs,
%we also define the ``best'' one and give an algoritm for its construction.

The most significant contribution of the paper is a sufficient
condition for equivalence of two NLFSRs before and after
the transformation. It is formulated and proved for the general case which
covers not only the equivalence between a Fibonacci and a Galois
NLFSRs, but only the equivalence between two Galois NLFSRs. 
%So, a possibility of transformation between two equivalent Galois
%NLFSRs is also considered.

%Content: 3 most significant contributions:
%- sufficient condition for existance of a non-linear recurrence
%describing the output sequence of an NLFSR
%- sufficient condition for the equivalence of two NLFSRs. MOST significant
%- definition of 

The paper is organized as follows. Section~\ref{prel} describes main
notions and definitions used in the sequel.  Section~\ref{rec_cond}
formulates a sufficient condition for existence of a non-linear
recurrence describing the output sequence of an NLFSR.
Section~\ref{suf_cond} presents a sufficient condition for the
equivalence of two NLFSRs. In Section~\ref{most_dec}, we define a
 Galois NLFSR which is unique for a given Fibonacci
NLFSR and show how to compute it. 
Section~\ref{conc} concludes the
paper and discusses open problems.

\section{Preliminaries} \label{prel}

In this section, we describe basic definitions and notation used in
the sequel.

The {\em algebraic normal form (ANF)} of a Boolean function $f:
\{0,1\}^n \rightarrow \{0,1\}$ is a polynomial in $GF(2)$ of type
\[
f(x_0, \ldots, x_{n-1}) = \sum_{i=0}^{2^n-1}  c_i \cdot 
x_0^{i_0} \cdot x_1^{i_1} \cdot \ldots \cdot x_{n-1}^{i_{n-1}},
\]
where $c_i \in \{0,1\}$ and $(i_0 i_1 \ldots i_{n-1})$ is the binary
expansion of $i$ with $i_{n-1}$ being the least significant bit.

The {\em dependence set} (or {\em support set}) of a Boolean function
$f$ is defined by
\[
dep(f) = \{i \ | \ f|_{x_i=0} \not = f|_{x_i=1}\},
\]
where $f|_{x_i=j} = f(x_0, \ldots, x_{i-1}, j, x_{i+1}, \ldots, x_{n-1})$
for $j \in \{0,1\}$.

Let $\alpha_{min}(f)$ ($\alpha_{max}(f)$) be the smallest (largest)
index of variables in $dep(f)$.

%The {\em Fibonacci} type of NLFSR, shown in Figure~\ref{nlfsr_f},
%consists of a number of {\em bits} numbered from right to left as $0,
%\ldots, n-1$ with feedback from each bit to the $n-1$th bit. At each
%clocking instance, the value of the bit $i$ is moved to the bit
%$i-1$. The value of the bit 0 becomes the output of the register. The
%new value of the bit $n-1$ is computed as some non-linear function of
%the previous values of bits $0, \ldots, n-1$, depending on the
%feedback function used.

%In the {\em Galois} type of NLFSR, shown in Figure~\ref{nlfsr_g}, each
%bit is updated according to its own feedback function. Thus, in contrast
%to the Fibonacci NLFSRs in which feedback is applied to the $n-1$th
%bit only, in a Galois NLFSR feedback is potentially applied to every
%bit.

%Throughout the paper, when we say {\em NLFSR} without mentioning the
%type, we mean that the statement is applicable to both, Fibonacci and
%Galois, types.

Let $f_i: \{0,1\}^n \rightarrow \{0,1\}$ be a feedback function of the
bit $i$, $i \in \{0,1,\ldots,n-1\}$, of an NLFSR.  All results in this
paper as derived for NLFSRs whose feedback functions are {\em
singular} functions of type
\begin{equation} \label{e_s}
f_i = x_{i+1} \oplus g_i(x_0,\ldots,x_{n-1}),
\end{equation}
where $g_i: \{0,1\}^{n-1} \rightarrow \{0,1\}$, $i+1 \not\in
dep(g_i)$, and the sign ``$+$'' is modulo $n$. Singularity guarantees
that the state transition graph of an NLFSR is ``branchless'',
i.e. that each state belongs to one of the state
cycles~\cite{Golomb_book}. 

Let $s_i(t)$ denote the value of the bit $i$ at time $t$.  The
sequence of states an $n$-bit NLFSR with the singular feedback
functions can be described by a system of $n$ non-linear equations of
type:
\begin{equation} \label{eq_sys}
\left\{
\begin{array}{l}
s_{n-1}(t) = s_0(t-1) \oplus g_{n-1}(s_1(t-1), s_2(t-1), \ldots, s_{n-1}(t-1)) \\[1mm]
s_{n-2}(t) = s_{n-1}(t-1) \oplus g_{n-2}(s_0(t-1), s_1(t-1), \ldots, s_{n-2}(t-1))\\
\ldots \\
s_0(t) = s_1(t-1) \oplus g_0(s_0(t-1), s_2(t-1), \ldots, s_{n-1}(t-1)).
\end{array}
\right.
\end{equation}

\section{A Condition for Existence of a Non-Linear Recurrence} \label{rec_cond}

In this section, we formulate a condition for existence of a
non-linear recurrence describing the output sequence of an NLFSR.
First, we introduce some definitions which are necessary for the
presentation of main results.

\begin{definition} \label{d_e}
Two NLFSRs are equivalent if there are initial states, possibly
different for each NLFSR, from which they generate the same output
sequences.
\end{definition}

\begin{definition} \label{d_f}
The feedback graph of an NLFSR has $n$ vertices
$v_0, \ldots, v_{n-1}$ representing the bits $0,\ldots,n-1$.
There is an edge from $v_i$ to $v_j$ if 
$i \in dep(f_j)$, $i,j \in \{0, 1,\ldots,n-1\}$.
\end{definition}

\begin{definition} \label{d_b}
The terminal bit of an $n$-bit NLFSR is the bit with the largest index
$i$ which satisfies the following condition: For all bits $j$ such
that $i > j \geq 0$, the feedback function $f_j$ is of type $f_j =
x_{j+1}$, $i,j \in \{0,1, \ldots,n-1\}$.
\end{definition}

\begin{definition} \label{d_s}
The operation substitution, denoted by $sub(v_i,v_j)$, is
defined for any vertex $v_i$ which has a unique predecessor $v_j$. The
substitution $sub(v_i,v_j)$ removes $v_i$ from the feedback graph and,
for each successor $v_k$ of $v_i$, replaces the edge $(v_i,v_k)$ by an
edge $(v_j,v_k)$, $i,j,k \in \{0,\ldots,n-1\}$.
\end{definition}

\begin{definition}
Given a feedback graph $G$, the reduced feedback graph of $G$ is a
graph obtained by subsequently applying the substitution to all
vertices of $G$ with the input degree 1.
\end{definition}

Since substitution merges a vertex with its unique predecessor, the
order of applying the substitution does not influence the resulting
reduced feedback graph, i.e. it is unique for a given $G$.

\begin{lemma} \label{l1}
If the feedback graph of an $n$-bit NLFSR can be reduced to a single
vertex $v_i$, $i \in \{0,1,\ldots,n-1\}$, then there exist a
non-linear recurrence describing the sequence of values of the bit $i$
of type
\begin{equation} \label{e_rec}
s_i(t) = 
\sum_{j=0}^{2^n-1}  (a_j \cdot \prod_{k=0}^{n-1} s^{j_k}(t-n+k)),
\end{equation}
where $a_j \in \{0,1\}$, $(j_0 j_1 \ldots j_{n-1})$ is the binary
expansion of $j$ with $j_{n-1}$ being the least significant bit, 
and $s^{j_k}(t-n+k)$ is defined as follows
\[
s^{j_k}(t-n+k) = 
\left\{
\begin{array}{ll}
s(t-n+k), & \mbox{for} \ i = 1, \\
1,	 & \mbox{for} \ i = 0. \\
\end{array}
\right.
\]
\end{lemma}
{\bf Proof:} Let $v_i$ be a vertex of the feedback
graph which has a unique predecessor $v_j$ and $m$ successors
$v_{k_1},\ldots,v_{k_m}$, $j, k_p \in \{0,1,\ldots,n-1\}$, $p \in
\{0,1,\ldots,m\}$. By Df.~\ref{d_f}, this implies that $s_i(t) =
s_j(t-1)$ and, for each $p$, $s_{k_p}(t)$ depends on $s_i(t-1)$.

The substitution $sub(v_i,v_j)$ is equivalent to replacing the
variable $s_i(t-1)$ in the equation of each $s_{k_p}(t)$ by
$s_j(t-2)$. This reduces the number of variables in the
equations~(\ref{eq_sys}) by one and reduces the number of
equations by one.

If the feedback graph of an NLFSR can be reduced to a single vertex,
say $v_r$, then the substitution can be applied $n-1$ times.  So, the
number of variables in the equations~(\ref{eq_sys}) can be reduced to a
single variable and the number of equations can be reduced to a single
equation. This equation corresponds to the non-linear recurrence
relation describing the sequence of states of the bit $r$ of the
NLFSR.
\begin{flushright}
$\Box$
\end{flushright}

%The above condition is sufficient, but not necessary. It is possible that
%we have an NLFSR which is not reducible, but a non-linear recurence exists
% Example
%\[
%\left\{
%\begin{array}{l}
%f_3 = x_0 \oplus 1 \oplus x_1 \oplus x_1 x_3, \\
%f_2 = x_3, \\
%f_1 = x_2 \oplus x_0, \\
%f_0 = x_1 .
%\end{array}
%\right.
%\]
%Reducidle to nodes 1 and 3 (with self-loops in both), 
%generates the following same output sequence as Fibonacci NLFSR 
%from Example 1, so recurrence exists for x1 (and x0).

%Furthermore, if the recurrence relation exists of for the bit $r$, it
%also exists for the remaining bits of NLFSR, since we can derive these
%recurrence relations by reversing the order in which the substitutions
%were applied to the feedback equations.

%\noindent {\bf only if}-part: Suppose that each bit of the NLFSR satisfies 
%some non-linear recurrence relation of order $n$, but the feedback
%graph can only be reduced to a set of vertices, say
%$v_{r_1},\ldots,v_{r_k}$.

%If the feedback graph cannot be reduced to less than $k$ vertices,
%then the system of $n$ feedback equations can only be reduced to $k$
%equations of $k$ variables.  Therefore, no recurrence relation of
%order $n$ exists for the bits $r_1,\ldots,r_k$, which contradicts the
%assumption.

%Let $i$ be the bit with the largest index which satisfy the condition:
%For all bits $j$ such that $i > j \geq 0$, $f_j = x_{j-1}$.
%Let $s_i(t)$ be the recurrence relation of $i$.

\noindent{\em Example 1:} 
As an example, consider a 4-bit Fibonacci NLFSR with the feedback
function $f_3 = x_0 \oplus x_1 \oplus x_2 \oplus x_1 x_3$. Its sequence
of states can be described by the following equations:
\[
\left\{
\begin{array}{l}
s_3(t) = s_0(t-1) \oplus s_1(t-1) \oplus s_2(t-1) \oplus s_1(t-1)s_3(t-1), \\s_2(t) = s_3(t-1), \\
s_1(t) = s_2(t-1), \\
s_0(t) = s_1(t-1).
\end{array}
\right.
\]
This NLFSR generates the following output sequence with the period 15:
\[
111011000101001 \ldots
\]

\begin{figure}[t!]
\begin{center}
\resizebox{0.85\columnwidth}{!} {\input{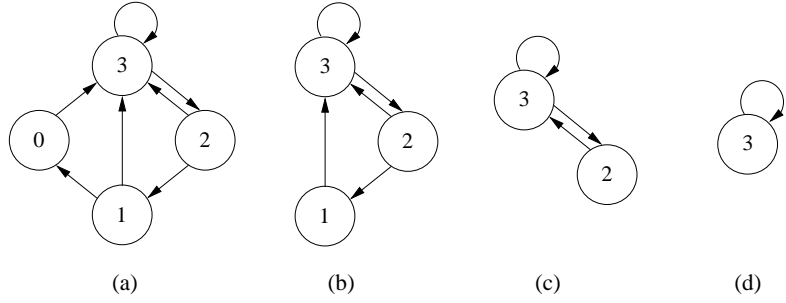}}
\caption{Reduction steps for the feedback graph of the Fibonacci
NLFSR from the example: (a) initial graph; (b) after $sub(v_0,v_1)$; 
(c) after $sub(v_1,v_2)$; (d) after $sub(v_2,v_3)$.}\label{f1}
\end{center}
\end{figure}

The feedback graph of this NLFSR is shown in Figure~\ref{f1}(a). It
can be reduced to a single vertex as follows:
\begin{enumerate}
\item $sub(v_0,v_1)$ reduces the graph to Figure~\ref{f1}(b).
This is equivalent to substituting $s_0(t)$ by $s_1(t-1)$ into the
equation of $s_3(t)$:
\[
s_3(t) = s_1(t-2) \oplus s_1(t-1) \oplus s_2(t-1) \oplus s_1(t-1)s_3(t-1).
\]
\item $sub(v_1,v_2)$ reduces the graph to Figure~\ref{f1}(c).
This is equivalent to substituting $s_1(t)$ by $s_2(t-1)$ into the
equation of $s_3(t)$:
\[
s_3(t) = s_2(t-3) \oplus s_2(t-2) \oplus s_2(t-1) \oplus s_2(t-2)s_3(t-1).
\]
\item $sub(v_2,v_3)$ reduces the graph to Figure~\ref{f1}(d).
This is equivalent to substituting $s_2(t)$ by $s_3(t-1)$ into the
equation of $s_3(t)$:
\[
s_3(t) = s_3(t-4) \oplus s_3(t-3) \oplus s_3(t-2) \oplus s_3(t-3)s_3(t-1).
\]
\end{enumerate}
This gives us a non-linear recurrence describing the sequence of values
of the bit 3. Since other bits repeat the content of the 3rd bit, the
recurrence is identical for all bits, and thus for the output of the NLFSR. 
%It generates the output sequence shown above.

It is easy to see that the feedback graph of a Fibonacci NLFSR can
always be reduced to a single vertex $v_{n-1}$. Therefore, for a
Fibonacci NLFSR, a non-linear recurrence of type~(\ref{e_rec}) always
exists. Its coefficients $a_i$, $i \in \{0,1,\ldots,2^n-1\}$, are
equal to the coefficients $c_i$ of the ANF of the feedback function
$f_{n-1}$.

For Galois NLFSRs, a non-linear recurrence of type~(\ref{e_rec}) may
or may not exist.  If it exists, it may be different for different
bits.

\noindent{\em Example 2:} 
As another example, consider a Galois NLFSR with the following 
feedback functions:
\[
\begin{array}{l}
f_3 = x_0 \oplus x_1 x_3, \\
f_2 = x_3, \\
f_1 = x_2, \\
f_0 = x_1 \oplus x_2 \oplus x_3.
\end{array}
\]
Its feedback graph can be reduced to the vertex $v_3$, giving us the following recurrence:
\[
s_3(t) = s_3(t-4) \oplus s_3(t-3) \oplus s_3(t-2) \oplus s_3(t-3)s_3(t-1).
\]
This recurrence is the same as the one of the Fibonacci NLFSR from the Example
1.  Bits 2 and 1 repeat the same recurrence as the bit 3, however, the
value of the bit 0 is the XOR of the bits 1, 2 and 3. Thus, its sequence of values differs
from the one of the 3rd bit. Therefore, the output sequence of this
Galois NLFSR, is different the output sequence of the Fibonacci NLFSR
from the Example 1.

%WRONG:%

%The recurrence for the bit 0 can be computed by adding recurrences for
%bits 1, 2 and 3:
%\[
%\begin{array}{rcl}
%s_0(t) & = & s_0(t-4) \oplus s_0(t-3) \oplus s_0(t-2) \oplus s_0(t-3)s_0(t-1) \\
%       & \oplus & s_0(t-1) \oplus s_0(t-4) \oplus s_0(t-3) \oplus s_0(t-4)s_0(t-2) \\
%       & \oplus & s_0(t-2) \oplus s_0(t-1) \oplus s_0(t-4) \oplus s_0(t-1)s_0(t-3)  \\
%       & = & s_0(t-4) \oplus s_0(t-4)s_0(t-2) \\
%\end{array}
%\]

\section{A Transformation from the Fibonacci to the Galois NLFSRs} \label{suf_cond}

In this section, we show how to transform a Fibonacci NLFSR into
an equivalent Galois NLFSR. 
%We show that such a decomposition always
%exist.  We also defined a type of ``most-perfectly uniform doecomposition.

%\begin{definition}
%The {\em splitting} transformation $f_a \rightarrow f_b$ 
%moves an ANF product-term $p = x_{i_1} x_{i_2} \ldots x_{i_r}$, $i_j
%\in \{0,1,\ldots,n-1\}$, $j \in \{1,2,\ldots,r-1\}$, $1 \leq r \leq n-1$, 
%from the feedback function $f_a$ to the feedback function $f_b$, $b < a$, $a,b \in
%\{0,1,\ldots,n-1\}$. The index of each variable in the moved product-term 
%is reduced by $a-b$, so that the product-term which is added to $f_b$
%is of type $x_{i_0-(a-b)} x_{i_1-(a-b)} \ldots x_{i_{n-1}-(a-b)}$
%where ``$-$'' is modulo $n$.

%f_a = f_a \oplus x_{i_1} x_{i_2} \ldots x_{i_r}
%f_b = f_b \oplus x_{i_0-(a-b)} x_{i_1-(a-b)} \ldots x_{i_{n-1}-(a-b)}
%\end{definition}

Let $P_f$ denote the set of all product-terms of the ANF of a function
$f: \{0,1\}^n \rightarrow \{0,1\}$.  Given an ANF product-term $p \in
P_f$, the notation $p_{-k}$ means that the index of each variable $x_i$
of $p$ is changed to $x_{i-k}$, where ``$-$'' is modulo $n$.

For example, if $n = 4$, and $p = x_0 x_1
x_3$ then
\[
p_{-1} = x_3 x_0 x_2, ~ p_{-2} = x_2 x_3 x_1, ~ p_{-3} = x_1 x_2 x_0. 
\]

\begin{definition} \label{d_sh}
The operation shifting, denoted by $f_a \stackrel{p}{\rightarrow}
f_b$, $p \in P_{f_a}$, $a,b \in \{0,1,\ldots,n-1\}$, $b < a$, removes
the product-term $p$ from the ANF of the function $f_a$ and adds the
product-term $p_{-(a-b)}$ to the ANF of the function $f_b$ .
\end{definition}

As we can see form the definition, shifting subtracts $(a-b)$ from the
index of each variable in the shifted product-term (modulo $n$). For
example, if initially
\[
\begin{array}{l}
f_3 = x_0 \oplus x_1 x_3 \\
f_2 = x_3
\end{array}
\]
then, after $f_3 \stackrel{x_1x_3}{\longrightarrow} f_2$, we get 
\[
\begin{array}{l}
f_3 = x_0 \\
f_2 = x_3 \oplus x_0 x_2.
\end{array}
\]
%The index of each variable in the shifted product-term is
%reduced by 1.

%\begin{definition}
%The shifting is {\em valid} if the initial and the
%transformed NLFSRs are equivalent.
%\end{definition}

%\begin{lemma}
%A shifting operation preserves the equivalence of the original and
%the transformed NLFSRs if it does not change the structure of the reduced
%feedback graph.
%\end{lemma}
%{\bf Proof:} Moving a term $x_{i_0} x_{i_1} \ldots x_{i_{r-1}}$ from
%$f_a$ to $f_b$, $b < a$, is equivalent to removing the term $s_{i_0}(t-1)
%s_{i_1}(t-1) \ldots s_{i_{r-1}}(t-1)$ from the feedback equation of
%$s_{a}(t)$ and adding the terms $s_{i_0}(t-1-(a-b)) s_{i_1}(t-1-(a-b))
%\ldots s_{i_{r-1}}(t-1-(a-b))$ feedback equation of $s_{b}(t)$.
%Since the number of clock units separaing the bits $a$ and $b$ is
%$a-b$, this transformation does not change the final recurrence
%relation.
%\begin{flushright}
%$\Box$
%\end{flushright}

%\begin{lemma}
%A shifting operation 
%\[
%(f_i, x_a x_b) \rightarrow (f_j, x_{a-(i-j)} x_{b-(i-j)})
%\]
%$i,j \in \{0,1,\ldots,n-1\}$, $i > j$,
%preserves the equivalence if for all $k$ such that 
%$n-1 > k \geq j$, $\alpha_{max}(f_k) \leq j$.
%\end{lemma}

%can be used to transform a Fibonacci
%NLFSR into an equivalent Galois NLFSR. It can also be used to
%transrotm an Galois NLFSR into another equivalent Galois NLFSR.

\begin{definition} \label{t_m} 
An $n$-bit NLFSR is uniform if: 
\begin{enumerate}
\item[(a)] all its feedback functions are of type~(\ref{e_s}), and 
\item[(b)] for all its bits $i$ such that $n-1 \geq i > \tau$, the following condition holds:
\begin{equation} \label{e1}
\alpha_{max}(g_i) \leq \tau,
\end{equation}
where $\tau$ is the terminal bit of the NLFSR, $\tau \in
\{0,1,\ldots,n-1\}$.
\end{enumerate}
\end{definition}

Note that any Fibonacci NLFSR is uniform. 

\begin{lemma} \label{l1n}
If an NLFSR is uniform, then its feedback graph can be reduced to a single vertex.
\end{lemma}
{\bf Proof:} Suppose that an NLFSR $N$ is uniform. We show that then we
can alway reduce the feedback graph of $N$ to the
vertex $v_{\tau}$ corresponding to the terminal bit $\tau$ of $N$.

By Df.~\ref{d_b}, for $i \in \{0,1,\ldots,\tau-1\}$, each vertex $v_i$ of
the feedback graph has input degree 1. So, for each $i \in
\{0,1,\ldots,\tau-1\}$, we can apply the substitution $sup(v_i,v_{i+1})$
to remove $v_i$ from the feedback graph, and, for each successor $v_k$
of $v_i$, to replace the the edge $(v_i,v_k)$ by an edge
$(v_{\tau},v_k)$. Therefore, by applying a sequence of substitutions
$sup(v_0,v_1)$, $sup(v_1,v_2)$, \ldots, $sup(v_{\tau-1},v_{\tau})$ we can
remove $v_0, v_1, \ldots, v_{\tau-1}$ from the feedback graph and change
the origin of all outgoing edges of $v_0, v_1, \ldots, v_{\tau-1}$ to
$v_{\tau}$.

Since the condition~(\ref{e1}) holds and the origin of all outgoing
edges of $v_0, v_1, \ldots, v_{\tau-1}$ is changed to $v_{\tau}$, each of the
vertices $v_i$ for $i \in \{\tau+1, \tau+2, \ldots, n-1\}$ has no more
than two incoming edges: one from $v_{i+1}$ and one from $v_{\tau}$. 
This implies that each of them has the output degree 1. 

Clearly, $v_{n-1}$ has only one incoming edge, from $v_{\tau}$. By applying
the substitution $sup(v_{n-1},v_{\tau})$, we can remove $v_{n-1}$ and
replace the edge $(v_{n-1},v_{n-2})$ by the edge $(v_{\tau},v_{n-2})$. This
make the input degree of $v_{n-2}$ one. Continuing similarly with the
sequence of substitutions $sup(v_{n-2},v_{\tau})$, \ldots,
$sup(v_{\tau+1},v_{\tau})$ we remove $v_{n-2}, \ldots, v_{\tau+1}$ and reduce the
graph to one vertex, $v_{\tau}$.
\begin{flushright}
$\Box$
\end{flushright}

The above condition is sufficient, but not necessary. For example, the
NLFSR from the Example 2 is not uniform, but it can be reduced to a
single vertex. 

The following theorem is the main result of the paper. It presents a
sufficient condition for equivalence of two NLFSRs. Note, that it
is formulated for shiftings on subfunctions $g_i$ of the singular
feedback functions $f_i$ (see the expression~\ref{e_s}), because the
variable $x_{i+1}$ should not be shifted in order to preserve the
register structure.

\begin{theorem} \label{t1}
%{\em (Suficcient condition)}
Given a uniform NLFSR, a shifting $g_a \stackrel{p}{\rightarrow} g_b$,
$a,b \in \{0,1,\ldots,n-1\}$, $b < a$, $P \subseteq P_{g_a}$,
preserves the equivalence if the transformed NLFSR is uniform as well.
\end{theorem}
{\bf Proof:} See Appendix.

%\begin{theorem} \label{t_m} {\em (Suficcient condition)}y
%Let $N$ be a Fibonacci NLFSR of type~(\ref{e_s}) and let $N^*$ be a
%Galois NLFSR obtained from $N$ by applying a sequence of shiftings 
%\[
%g_{a_1} \stackrel{P_1}{\longrightarrow} g_{b_1}, \ g_{a_2} \stackrel{P_2}{\longrightarrow} g_{b_2}, \ldots, \ g_{a_k} \stackrel{P_k}{\longrightarrow} g_{b_k},
%\]
%for some $a_i,b_i \in \{0,1,\ldots,n-1\}$, $b_i < a_i$, and $P_i \subseteq G_{a_i}$,
%$i = \{1,2,\ldots,k\}$.

%Then $N^*$ is equivalent to $N$ if, for all bits $j$ of $N^*$ such that
%$n-1 \leq j < \tau$ the following condition holds:
%\begin{equation} \label{e1}
%\alpha_{max}(g_j) \leq \alpha_{max}(g_{\tau}),
%\end{equation}
%where $\tau$ is the terminal bit of $N^*$, $\tau \in
%\{0,1,\ldots,n-1\}$.
%\end{theorem}

The condition of the Theorem~\ref{t1} is sufficient, but not necessary. 
For example, the following NLFSR can be obtained from
the NLFSR from the Example 1 by applying the shifting
$f_3 \stackrel{x_1x_3}{\longrightarrow} f_0$,
$f_3 \stackrel{x_1}{\longrightarrow} f_1$
 and
$f_3 \stackrel{x_2}{\longrightarrow} f_1$:
\[
\begin{array}{l}
f_3 = x_0, \\
f_2 = x_3, \\
f_1 = x_2 \oplus x_0 \oplus x_3, \\
f_0 = x_1 \oplus x_0 x_2.
\end{array}
\]
This NLFSR is not uniform, however, it is equivalent to the NLFSR from the Example 1.
% its feedbacj graph reducec to vertex 0

Next, we formulate a condition which should be satisfied in order to
obtain a uniform NLFSR after shifting.

\begin{theorem} \label{t2} 
%{\em (Necessary condition)} 
Given a uniform NLFSR $N$, an NLFSR obtained from $N$ by a
shifting $g_a \stackrel{p}{\rightarrow} g_b$, $a,b \in \{0,1,\ldots,n-1\}$, $b < a$, $P \subseteq P_{g_a}$, is uniform 
only if
\begin{equation} \label{e3}
b \geq a - \alpha_{min}(p).
\end{equation}
\end{theorem}
{\bf Proof:} If $b < a - \alpha_{min}(p)$, then $\alpha_{min}(p) <
a-b$.  Therefore, after the shifting $g_a \stackrel{p}{\rightarrow}
g_b$, $\alpha_{min}(p)$ becomes $\alpha_{min}(p) + n-(a-b) =
\alpha_{min}(p) + b+(n-a)$. By Df.~\ref{d_sh}, $b < a$, thus $a$ is
always greater than 0. So, for any $a \in \{1,2,\ldots,n-1\}$, after
shifting the feedback function $g_b$ contains a product-term whose
index is greater than $b$ by $n-a$. Since the terminal bit of the
NLFSR is smaller or equal to $b$, the condition~(\ref{e1}) of
Df.~\ref{t_m} is violated.
\begin{flushright}
$\Box$
\end{flushright}

%The above Theorem shows that, for a given $n$-bit Fibonacci NLFSR, the
%smallest possible terminal bit in an equivalent Galois NLFSR is
%$n-1-\alpha_{min}(G_{n-1})$. WRONG?

Often an equivalent Galois NLFSR can be obtained from a Fibonacci
NLFSR by shifting product-terms one-by-one.  Sometimes, however, more
than one product-term has to be shifted in order to preserve the
equivalence.  For example, if the feedback function $g_{n-1}$ has more
than one product-term containing the variable $x_{n-1}$, then all such
product-terms have to be shifted. The Lemma below describes two cases
in which the product-terms can be shifted one-by-one.

\begin{lemma} \label{t3}
Given a uniform NLFSR with the terminal bit $\tau$ and a sifting $g_a
\stackrel{p}{\rightarrow} g_b$, $a,b \in \{0,1,\ldots,n-1\}$, $b < a$, $P \subseteq P_{g_a}$,
%$a,b \in \{0,1,\ldots,n-1\}$, $b < a$, $P \subseteq P_{g_a}$, 
the following holds:
%which satisfies the condition~(\ref{e3}).
\begin{enumerate}
\item[(a)]If $b \geq \tau$, then $g_a \stackrel{p}{\rightarrow} g_b$
preserves the equivalence for any $p \in P_{g_a}$ which satisfies the condition~(\ref{e3}).
\item[(b)] If $b < \tau$ and $\alpha_{max}(g_i) \leq b$ for all $i \in
\{n-1,n-2,\ldots,b\}$, then $g_a \stackrel{p}{\rightarrow} g_b$
preserves the equivalence for any $p \in P_{g_a}$ which satisfies the condition~(\ref{e3}).
%\item[(c)] If $b < \tau$ and $\alpha_{max}(g_i) > b$ for some $i \in
%\{0,1,\ldots,n-1\}$, then the original and transformed NLFSRs are equivalent
%if and only if, for each $g_i$ such that $\alpha_{max}(g_i) > b$, all
%product-terms containing a variable with an index greater than $b$ are
%moved to $f_b$, and each shiftiing $g_i \stackrel{p}{\rightarrow} g_b$
%satisfies the condition~(\ref{e3}).
\end{enumerate}
\end{lemma}
{\bf Proof:} {\bf Case (a):} By Df.~\ref{d_sh}, after the shifting
$\alpha_{min}(p)$ becomes $\alpha_{min}(p)-(a-b)$. Since the condition
(\ref{e3}) is satisfied, $\alpha_{min}(p) \geq a-b$, i.e. after
shifting the indexes of variables of $p$ are reduced by some value
between 1 and $\alpha_{min}(p)$. Therefore, after the shifting, none
of the product-terms of $p$ violates the condition~(\ref{e1}). Since
the initial NLFSR is uniform and the terminal bit is not changed, the
transformed NLFSR is uniform as well, and therefore, by
Theorem~\ref{t1}, the equivalence is preserved.

\noindent {\bf Case (b):} Similarly to the case (a) we can show that, 
after the shifting, none of the product-terms of $p$ violates the
condition~(\ref{e1}). Since $\alpha_{max}(g_i) \leq b$ for all $i$ by
assumption, the transformed NLFSR is uniform and therefore, by
Theorem~\ref{t1}, the equivalence is preserved.

%\noindent{\bf case (c):} 
%\noindent {\bf if}-part: 
%Suppose that, for each $g_i$ such that $\alpha_{max}(g_i) > b$, all
%product-terms containing a variable with an index greater than $b$ are
%moved to $f_b$, and each shiftiing satisfies the condition~(\ref{e3}).
%Since the original NLFSR is uniform, the largest value of
%$\alpha_{max}(g_i)$ is $\tau$. Since $a \leq \tau$ and $b < a$, 
%we get $\tau - (a-b) \leq b$, i.e. the largest index of variables
%in each shifted product-term is smaller or equal than $b$.

%On the other hand, similarly the the case (a) we can show that the
%smalest index of variables in each shifted product-term is greater or
%equal than 0. Therefore, the transformed NLFSR is uniform, by
%Theorem~\ref{t1}, the shifting is valid.

%\noindnent{\bf only if}-part: If some product-term $P$ such that
%$\alpha_{max}(p) > b$ of some $g_i$ is not shifted, then the
%condition~(\ref{e1}) is violated since $i > b$ and $g_i$ contains a
%variable with the index greater than $b$.
\begin{flushright}
$\Box$
\end{flushright}

The above Lemma implies that, for any Fibonacci NLFSR, shifting can
always reduce the index of the initial terminal bit $n-1$ at least by 1. It
reduces the index of the terminal bit exactly by 1 if $g_{n-1}$ of
the Fibonacci NLFSR contains a product with $\alpha_{max}(g_i) = n-1$
and $\alpha_{min}(g_{n-1}) = 1$.  The smaller the difference between
$\alpha_{max}(g_{n-1})$ and $\alpha_{min}(g_{n-1})$, the more the
index of the initial terminal bit can be reduced.

\section{Fully Shifted Galois NLFSRs} \label{most_dec}

Usually, there are multiple ways to transform a Fibonacci NLFSR into a
Galois NLFSR. Next, we define a ``fully shifted'' Galois NLFSR which is
unique for a given Fibonacci NLFSR and show how to compute it. 

\begin{definition}
An NLFSR is fully shifted if no product-term of any function
$g_i$ can be shifted to a function $g_j$ with the index $j < i$
without violating the condition~(\ref{e1}), $i,j \in
\{0,1,\ldots,n-1\}$.
\end{definition}

%We are emphasizing this specific case for two reasons: 
%\begin{enumerate}
%\item 
In the linear case, a fully shifted NLFSR reduces to 
a Galois LFSR, i.e. it is a generalization of the Galois LFSR.
Note that this is not the case for NLFSRs which are not
fully shifted.
%\item In a fully shifted NLFSR, the number of feedback variables, i.e.
%the size of the set
%\[
%{\displaystyle \bigcup_{i=0}^{n-1} dep(g_i)} 
%\]
%is minimized (assuming that the equivalence is defined as in Df.~\ref{d_e}).
%\end{enumerate}

{\em Algorithm 1:} Given a uniform $n$-bit Fibonacci NLFSR $N$, the
fully shifted Galois NLFSR $\hat{N}$ which is equivalent to $N$ is obtained
as follows.

First, the terminal bit $\tau$ of $\hat{N}$ is computed as:
\begin{equation} \label{d_tau}
\begin{array}{lccl}
\tau & = & max & (\alpha_{max}(p)-\alpha_{min}(p)), \\
     & & \forall p \in P_{g_{n-1}} & \\
     & & \mbox{with} |p| > 1 & 
\end{array}
\end{equation}
where $|p|$ denotes the number of variables in the product-term $p$.
%and $P_{g_{n-1}}$ is the set of all product-terms of the ANF of the
%feedback function $g_{n-1}$ of $N$.

Then, each product-term $p \in P_{g_{n-1}}$ with $\alpha_{min}(p) \leq
(n-1)-\tau$ is shifted to $g_{n-1-\alpha_{min}(p)}$:
\[
g_{n-1} \stackrel{p}{\longrightarrow} g_{n-1-\alpha_{min}(p)}.
\]
and each product-term $p \in P_{g_{n-1}}$ with $\alpha_{min}(p) > (n-1)-\tau$ is shifted 
to $g_{\tau}$:
\[
g_{n-1} \stackrel{p}{\longrightarrow} g_{\tau}.
\]

\begin{theorem}
Algorithm 1 correctly computes the fully shifted Galois NLFSR for
a given Fibonacci NLFSR.
\end{theorem}
{\bf Proof:} For each product $p$ such that $\alpha_{min}(p) \leq
(n-1)-\tau$, the indexes are reduces by $\alpha_{min}(p)$.  So, after the shifting, the
smallest index becomes 0 and the largest becomes
$\alpha_{max}(p)-\alpha_{min}(p)$.  By equation~(\ref{d_tau}),
$\alpha_{max}(p)-\alpha_{min}(p) \leq \tau$.

For each product $p$ such that $\alpha_{min}(p) > (n-1)-\tau$, 
the indexes are reduces by $(n-1)-\tau$.  
Since $\alpha_{min}(p) < \alpha_{max}(p) \leq n-1$, 
the largest index after the shifting is 
$0 < \alpha_{max}(p)-((n-1)-\tau) \leq \tau$.
Since $(n-1)-\tau < \alpha_{min}(p) < \alpha_{max}(p)$, the smallest index
after the shifting is $0 < \alpha_{min}(p)-((n-1)-\tau) < \tau$.

So, the transformed NLFSR $\hat{N}$ is uniform and therefore, by
Theorem~\ref{t1}, two NLFSRs are equivalent. It remains to prove that
$\hat{N}$ is fully shifted.

By Df~\ref{d_sh}, index of each variable of $p$ is reduced by
$\alpha_{min}(p)$ after the shifting. Therefore, for each product-term $p
\in P_{g_{n-1}}$ such that $\alpha_{min}(p) \leq \tau$, $p$ after the shifting
contains a variable $x_0$.  If $p$ is shifted further from
$g_{n-1-\alpha_{min}(p)}$ to $g_{n-1-\alpha_{min}(p)-i}$ for some $1
\leq i \leq n-1-\alpha_{min}(p)$, the index of $x_0$
increases to $n-i$. For every value of $i$ in the range $1
\leq i \leq n-1-\alpha_{min}(p)$, $n-i > n-1-\alpha_{min}(p)$,
so the condition~(\ref{e1}) is violated and the resulting NLFSR is not
equivalent to the initial Fibonacci NLFSR.

Each product-term $p \in P_{g_{n-1}}$ such that $\alpha_{min}(p) > \tau$
is shifted to the terminal bit $\tau$. If $p$ is shifted to some $i <
\tau$, then, according to the equation~(\ref{d_tau}),
there is a product-term $p^*$ which has $\alpha_{max}(p^*) > i$ after
shifting. Thus, the condition~(\ref{e1}) is violated and the resulting
NLFSR is not equivalent to the initial Fibonacci NLFSR.
\begin{flushright}
$\Box$
\end{flushright}

%\noindent{\em Example 3:} 
%The fully shifted Galois NLFSR for the Fibonacci NLFSR $f_3 = x_0 \oplus
%x_1 \oplus x_2 \oplus x_1 x_3$ from the Example 1 is:
%\[
%\begin{array}{l}
%f_3 = x_0, \\
%f_2 = x_3 \oplus x_0 \oplus x_0 x_2, \\
%f_1 = x_2 \oplus x_0, \\
%f_0 = x_1. 
%\end{array}
%\]

\noindent{\em Example 4:} As an example, consider the following
32-bit Fibonacci NLFSR which is used in the NLFSR-based stream cipher
from~\cite{GaGK06}:
\[
\begin{array}{ll}
f_{31} & = \ x_0 \oplus x_2 \oplus x_6 \oplus x_7 \oplus x_{12} \oplus x_{17} \oplus x_{20} \oplus x_{27} \oplus x_{30} \oplus x_3 x_9 \oplus \ x_{12} x_{15} \oplus x_4 x_5 x_{16}
\end{array}
\]

Its corresponding fully shifted Galois NLFSR has the terminal bit
$\tau = 12$ and the following feedback functions:
\[
\begin{array}{l}
f_{31} = x_0 \\
f_{29} = x_{30} \oplus x_0 \\
f_{28} = x_{29} \oplus  x_0 x_6  \\
f_{27} = x_{28} \oplus  x_0 x_1 x_{12} \\
f_{25} = x_{26} \oplus x_0 \\
f_{24} = x_{25} \oplus x_0 \\
f_{19} = x_{20} \oplus x_0 \oplus x_0 x_3 \\
f_{14} = x_{15} \oplus x_0 \\
f_{12} = x_{13} \oplus  x_1 \oplus  x_8 \oplus x_{11} \\
%f_{11} = x_{12} \oplus x_0 \\
%f_4 = x_5 \oplus x_0 \\
%f_1 = x_2 \oplus x_0 \\
\end{array}
\]
The functions which are omitted are of type $f_i = f_{i+1}$. 
This NLFSR has 7 feedback variables: $x_0, x_1, x_3, x_6,
x_8, x_{11}$ and $x_{12}$, while the Fibonacci NLFSR has 15 feedback variables.

We can further reduce the depth of circuits implementing feedback functions
and the number of feedback variables as follows:
\[
\begin{array}{l}
f_{31} = x_0 \\
f_{29} = x_{30} \oplus x_0 \\
f_{28} = x_{29} \oplus  x_0 x_6  \\
f_{27} = x_{28} \oplus  x_0 x_1 x_{12} \\
f_{25} = x_{26} \oplus x_0 \\
f_{24} = x_{25} \oplus x_0 \\
f_{20} = x_{21} \oplus x_1 x_4 \\
f_{19} = x_{20} \oplus x_0  \\
f_{16} = x_{17} \oplus  x_{12} \\
f_{14} = x_{15} \oplus x_0 \\
f_{13} = x_{14} \oplus  x_{12} \\
f_{12} = x_{13} \oplus  x_1  \\
\end{array}
\]
This NLFSR has 5 feedback variables: $x_0, x_1, x_4, x_6$ and $x_{12}$.

\section{Conclusion} \label{conc}

In this paper, we show how to transform a Fibonacci NLFSR into the
Galois configuration.
%, and identify the unique Galois NLFSR which minimizes the
%number of feedback variables.
% which seems to provide the
%best advantage in speed with no penalty in cost. 

%In the future work, we plan to investigate whether it would be of
%advantage to weaken of the definition of equivalence of two NLFSR from
%the equivalence of output sequences to the equivalence of sequences of
%any two bits. For example, in this case NLFSRs from the Examples 1
%and 2 would be considered weakly equivalent.

The most important open problem is finding an algorithm for
constructing NLFSRs with a guaranteed long period. This problem is
hard because there seems to be no simple algebraic theory supporting
it. Specifically, primitive generator polynomials for LFSR have no
analog in the nonlinear case.

\bibliographystyle{ieeetr}
\bibliography{bib}

\section{Appendix: Proof of the Theorem~\ref{t1}} 
%{\bf Appendix: Proof of the Theorem~\ref{t1}:} 

Suppose that the transformed NLFSR is uniform. Then, by
Lemma~\ref{l1n}, its feedback graph can be reduced to the vertex $v_{b}$
corresponding to the terminal bit $b$ of the transformed NLFSR after the
shifting $g_a \stackrel{p}{\rightarrow} g_b$.  So, by Lemma~\ref{l1},
there exists a non-linear recurrence describing the sequence of values
of the bit $b$. It remains to prove that this recurrence is the same
as the one of the initial NLFSR.

It is sufficient to consider the case when the shifting $g_a
\stackrel{p}{\rightarrow} g_b$ moves a product-term of type
$x_k x_a$ for some $k < a$. For product-terms with more variables or
the product-term without $x_a$ the proof is similar.

If the shifted product is $x_k x_a$, then the function $g_a$ can be
represented as $g_a = g_a^* \oplus x_k x_a$, where $g_a^* = g_a \oplus
x_k x_a$.  So, the NLFSR before the shifting can be represented by the
following system of equations:
\[
\left\{
\begin{array}{l}
s_{n-1}(t) = s_0(t-1) \oplus g_{n-1}(s_0(t-1),s_1(t-1),\ldots,s_b(t-1)) \\
\ldots \\
s_a(t) = s_{a+1}(t-1) \oplus g^*_a(s_0(t-1),s_1(t-1),\ldots,s_b(t-1)) \oplus s_k(t-1) s_a(t-1)\\[1mm]
s_{a-1}(t) = s_a(t-1) \\
\ldots \\
s_0(t) = s_1(t-1) \\
\end{array}
\right.
\]
Since $i+1 \not\in dep(g_i)$ for $i \in \{0,1,\ldots,n-1\}$, each
$g_i$ does not depends of $s_{i+1}(t-1)$. However, we keep this
redundant term in the equations in order to be able to later introduce
the same abbreviations for all $g_i$.

Note, that each of $g_{n-1}, g_{n-2}, \ldots, g^*_a$ depends on variables
with indexes smaller or equal than $b$ only since, by assumption, the
condition~(\ref{e1}) holds after the shifting.

A substitution $sub(v_i,v_{i+1})$ is equivalent to replacing the
variable $s_i(t-1)$ in the equation of each successor of $v_i$ by
$s_{i+1}(t-2)$. After the sequence of $a$ substitutions
$sup(v_0,v_1),\ldots,sup(v_{a-1},v_a)$, each $s_i(t-1)$ gets replaced
by $s_a(t-1-(a-i))$, so the above equations reduce to:
\[
\left\{
\begin{array}{ll}
s_{n-1}(t) & = \ s_a(t-a-1) \oplus \ g_{n-1}(s_a(t-a-1),s_a(t-a),\ldots,s_a(t-1-(a-b))) \\
\ldots \\
s_a(t) & = \ s_{a+1}(t-1) \oplus  s_a(t-1-a+k) s_a(t-1) \\[1mm]
	& \oplus \ g^*_a(s_a(t-a-1),s_a(t-a),\ldots,s_a(t-1-(a-b))) \\[1mm]
	 
\end{array}
\right.
\]

To shorten the expressions, let us introduce an abbreviation
$\tilde{s}_a := (s_a(t-a-1),s_a(t-a),\ldots,s_a(t-1-(a-b)))$ and let
the notation $\tilde{s}_a(i)$ mean that each element $s_a(x)$
$\tilde{s}_a$ of is replaced by $s_a(x+i)$. For example,
$\tilde{s}_a(-1) = (s_a(t-a-2),s_a(t-a-1),\ldots,s_a(t-2-(a-b)))$.
Then, the above equations can be re-written us:

\[
\left\{
\begin{array}{l}
s_{n-1}(t) = s_a(t-a-1) \oplus g_{n-1}(\tilde{s}_a) \\
\ldots \\
s_a(t) = s_{a+1}(t-1) \oplus g^*_a(\tilde{s}_a) \oplus s_a(t-1-a+i) s_a(t-1) \\
\end{array}
\right.
\]

After a sequence of $n-a-1$ substitutions $sub(v_{n-1},v_{n-2}),\ldots,sub(v_{a+1},v_a)$,
we get a non-linear recurrence describing the sequence of values of the bit $a$:
\[
\begin{array}{ll}
s_a(t) \ & = \ s_a(t-n) \oplus g_{n-1}(\tilde{s}_a(-n+a+1)) \oplus g_{n-2}(\tilde{s}_a(-n+a)) \\[1mm]
         & \oplus \ \ldots \oplus g^*_a(\tilde{s}_a)+s_a(t-1-a+i) s_a(t-1)
\end{array}
\]

After expanding the abbreviation $\tilde{s}_a$, the above recurrence becomes:

\begin{equation} \label{rec_a}
\begin{array}{ll}
s_a(t) \ & = \ s_a(t-n) \\[1mm]
 	 & \oplus \ g_{n-1}(s_a(t-n),s_a(t-n+1),\ldots,s_a(t-n+b)) \\[1mm]
 	 & \oplus \ g_{n-2}(s_a(t-n-1),s_a(t-n),\ldots,s_a(t-n+b-1)) \\
	 & \ldots \\
         & \oplus \ g^*_a(s_a(t-a-1),s_a(t-a),\ldots,s_a(t-1-a+b)) \\[1mm]
	 & \oplus \ s_a(t-1-a+i) s_a(t-1)
\end{array}
\end{equation}

On the other hand, the NLFSR after the shifting can be represented by the
following system of equations:
\[
\left\{
\begin{array}{l}
s_{n-1}(t) = s_0(t-1) \oplus g_{n-1}(s_0(t-1),s_1(t-1),\ldots,s_b(t-1)) \\
\ldots \\
s_a(t) = s_{a+1}(t-1) \oplus g_a(s_0(t-1),s_1(t-1),\ldots,s_b(t-1)) \\[1mm]
s_{a-1}(t) = s_a(t-1) \\
\ldots \\
s_b(t) = s_{b+1}(t-1) \oplus s_{i-(a-b)}(t-1) s_b(t-1) \\
\ldots \\
s_0(t) = s_1(t-1) \\
\end{array}
\right.
\]

After the sequence of $b$ substitutions $sup(v_0,v_1),\ldots,sup(v_{b-1},v_b)$ we get:
\[
\left\{
\begin{array}{l}
s_{n-1}(t) = s_b(t-b-1) \oplus g_{n-1}(s_b(t-b-1),s_1(t-b),\ldots,s_b(t-1)) \\
\ldots \\
\ldots \\
s_a(t) = s_{a+1}(t-1) \oplus g^*_a(s_b(t-b-1),s_1(t-b),\ldots,s_b(t-1)) \\[1mm]
s_{a-1}(t) = s_a(t-1) \\
\ldots \\
s_b(t) = s_{b+1}(t-1) \oplus s_b(t-1+i-a) s_b(t-1)
\end{array}
\right.
\]

Introducing an abbreviation $\tilde{s}_b := (s_b(t-b-1),s_b(t-b),\ldots,s_b(t-1))$
we can re-write the above equations us:
\[
\left\{
\begin{array}{l}
s_{n-1}(t) = s_b(t-b-1) \oplus g_{n-1}(\tilde{s}_b) \\
\ldots \\
s_a(t) = s_{a+1}(t-1) \oplus g^*_a(\tilde{s}_b) \\[1mm]
s_{a-1}(t) = s_a(t-1) \\
\ldots \\
s_b(t) = s_{b+1}(t-1) \oplus s_b(t-1+i-a) s_b(t-1)
\end{array}
\right.
\]

After the sequence of $n-b-1$ substitutions $sub(v_{n-1},v_{n-2}),\ldots,sub(v_{b+1},v_b)$, 
we get a non-linear recurrence describing the sequence of values of the bit $b$:
\[
\begin{array}{ll}
s_b(t) \ & = \ s_b(t-n) \oplus g_{n-1}(\tilde{s}_b(-n+b+1)) \oplus g_{n-2}(\tilde{s}_b(-n+b)) \\[1mm]
         & \ \oplus \ldots \oplus  g^*_b(\tilde{s}_b(-(a-b)) \oplus s_b(t-1+i-a) s_b(t-1)
\end{array}
\]

After expanding the abbreviation $\tilde{s}_b$, the above recurrence becomes:
\begin{equation} \label{rec_b}
\begin{array}{ll}
s_b(t) \ & = \ s_b(t-n) \\[1mm]
         & \oplus \ g_{n-1}(s_b(t-n),s_b(t-n+1),\ldots,s_b(t-n+b)) \\[1mm]
         & \oplus \ g_{n-2}(s_b(t-n-1),s_b(t-n),\ldots,s_b(t-n+b-1)) \\
	 & \ldots \\
         & \oplus \ g^*_b(s_b(t-a-1),s_b(t-a),\ldots,s_b(t-1-a+b)) \\[1mm]
         & \oplus \ s_b(t-1-a+i) s_b(t-1)
\end{array}
\end{equation}

The non-linear recurrences~(\ref{rec_a}) and (\ref{rec_b}) are the same,
so two NLFSRs are equivalent.
\begin{flushright}
$\Box$
\end{flushright}

\end{document}